\documentclass[10pt]{mpi08}
\usepackage{graphicx}
\usepackage{cite,./mcite}

\begin{document} \title{MPI08 \\ QCD Mini-jet contribution to the total
cross section}
\author{A. Achilli$^1$\protect\footnote{\ \ speaker},
R. Godbole$^2$, A.Grau$^3$, G. Pancheri$^4$, Y.N. Srivastava$^1$}
\institute{$^1$INFN and Physics Department, University of Perugia, I-06123 Perugia, Italy
\\
$^2$ Centre for High Energy Physics, Indian Institute of Science,Bangalore, 560012, India
\\
$^3$Departamento de F\'\i sica Te\'orica y del Cosmos,
Universidad de Granada, 18071 Granada, Spain
\\
$^4$INFN Frascati National Laboratories, I-00044 Frascati, Italy}
\maketitle
\begin{abstract}
We present the predictions of a model for proton-proton total
cross-section at LHC. It takes into account both hard partonic
processes and soft gluon emission effects to describe the proper
high energy behavior and to respect the Froissart bound.
\end{abstract}

\section{Introduction}
A reliable prediction of the total proton-proton cross section is
fundamental to know which will be the underlying activity at the
LHC and for new discoveries in physics from the LHC data. In this
article, we shall describe a model \cite{lastPLB} \cite{ourmodel}
for the hadronic total cross section based on QCD minijet
formalism. The model  includes a resummation of soft gluon
radiation  which  is necessary to tame the fast high-energy rise
typical of a purely perturbative minijet model.  It is called the
BN model from the Bloch and Nordsiek discussion  of the infrared
catastrophe in QED. In the first section, results are presented
concerning the behavior of the QCD minijet cross section. It will
then be explained how this term is included into an eikonal
formalism where infrared soft gluon emission effects are added.
The last section is devoted to
the link between  the total cross-section asymptotic high energy
behavior predicted by our model  and the model parameters. This
relation also shows that our prediction is in agreement with the
limit imposed by the Froissart bound.

\section{Mini-jet cross section}
\begin{figure}[htb!]
  \begin{center}
  \includegraphics[height=3.9in]{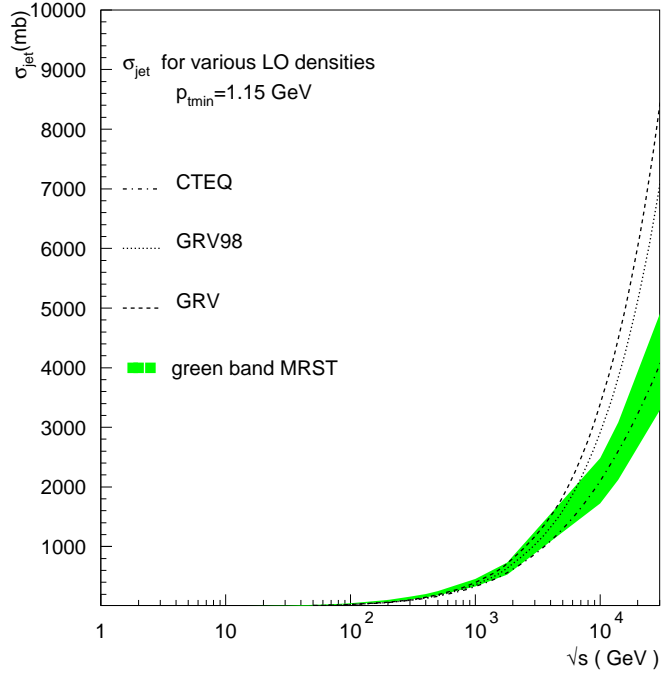}
  \caption{minijet cross section for different input parton densities.}
  \label{fig:1}
  \end{center}
\end{figure}

Hard processes involving high-energy partonic collisions drive the
rise of the total cross section \cite{cline}. These jet-producing collisions
are typical perturbative processes and we can describe them
through the usual QCD expression:
\begin{equation}
 \sigma^{AB}_{\rm jet} (s,p_{tmin})=
\int_{p_{tmin}}^{\sqrt{s/2}} d p_t \int_{4
p_t^2/s}^1 d x_1  \int_{4 p_t^2/(x_1 s)}^1 d x_2 \times
\sum_{i,j,k,l}
f_{i|A}(x_1,p_t^2) f_{j|B}(x_2,p_t^2)
  \frac { d \hat{\sigma}_{ij}^{ kl}(\hat{s})} {d p_t},
  \label{eq1}
  \end{equation}
with $A,B = p, \bar p$. This expression depends on the parameter
$p_{tmin}$ which represents the minimum transverse momentum of the
scattered partons for which one allows a perturbative QCD
treatment.
Its value is usually
around $\approx 1-2$ GeV and it distinguishes hard processes (that
are processes for which a perturbative approach is used) from the
soft ones that dominate at low energy, typically for $\sqrt{s}\le 10\div 20 \  GeV$, i.e, well before the cross-section starts rising. The Minijet expression also
depends on the DGLAP evoluted Partonic Densities Functions $f_{i|A}$ for
which there exist in the literature different LO parameterizations(GRV, MRST, CTEQ \cite{densities}). We obtain an
asymptotic growth of $\sigma_{jet}$ with energy as a power of $s$. As shown in figure \ref{fig:1}, the value of the exponent depends
on the PDF used and one has
\[
\ \sigma_{jet}^{GRV}\approx s^{0.4} \ \ \ \
\sigma_{jet}^{MRST}\approx s^{0.3} \ \ \ \
\sigma_{jet}^{CTEQ}\approx s^{0.3}.
\]
This result can be derived  by considering the relevant
contribution to the integral in (\ref{eq1}) in the
$\sqrt{s}>>p_{tmin}$ limit. In this limit, the major contribution
comes from the small fractions of momentum carried by the
colliding gluons
with  $x_{1,2}<<1$. In this limit we know that the relevant PDF's
behave approximately like powers of the momentum fraction $x^{-J}$
with $J \sim 1.3$ \cite{Lomatch}. From the previous consideration
and noting that $\frac { d \hat{\sigma}_{ij}^{ kl}(\hat{s})} {d
p_t} \propto \frac{1}{p_t ^3}$ we obtain from (\ref{eq1}) the
following asymptotic high-energy expression for $\sigma_{jet}$:

\begin{equation}
\sigma _{jet}  \propto \frac{1}{{p_{t\min }^2 }}\left[
{\frac{s}{{4p_{t\min }^2 }}} \right]^{J - 1}.
\end{equation}
The dominant term is just a power of $s$ and the estimate obtained
for the exponent $\epsilon=J-1\sim0.3$ is in agreement with our
previous results.
We now need to understand how to incorporate into a model for the total cross section this very fast rise at very high energy, which is present
in the perturbative regime.
Firstly it
is important to note that $\sigma_{jet}$ is an inclusive cross
section and therefore contains in itself a multiplicity factor,
linked to the average number $< n >$ of partonic collisions that
take place during the hadronic scattering. We can approximate the energy driving term at high energy
\cite{durand} $< n >$ as
\begin{equation}
 < n > \approx \sigma_{jet} \cdot A ,
\label{mult}
\end{equation}
where $A$ is a function representing the overlap between the two hadrons.


Now we can derive an expression for the total cross section as a
function of $<n>$. Assuming that the number of partonic collisions
follows  a Poisson distribution, since each interaction is
independent from the other, the probability of having $k$ partonic
collisions is:
\begin{equation}
 P(k, < n > ) = \frac{{ < n > ^k e^{ -  < n > } }}{{k!}}.
\end{equation}
The average number of partonic collisions should depend on the
energy and on the impact parameter $b$ relative to the hadronic
process $< n > \equiv < n(b,s) >$. From the previous expression it
is possible to obtain the inelastic hadronic cross section:
\begin{equation}
 \sigma _{inelastic}  = \int {d^2 b\sum\limits_{k = 1} {P(k, < n(b,s) > )} }  = \int {d^2 b\left[ {1 - e^{ -  < n(b,s) > } }
 \right]} ,
\end{equation}
which is the usual eikonal expression if we consider  the link
between $< n(b,s) >$ and the eikonal $\chi (b,s)$:
\begin{equation}
  < n(b,s) >  = 2{\mathop{\rm Im}\nolimits} \chi (b,s).
\end{equation}

\section{Eikonal model}

The eikonal representation allows to implement multiple parton
scattering and to restore a finite size of the interaction.
Neglecting the real part of the eikonal function,  an acceptable
approximation in the high energy limit,  the expression for the
total cross section is
\begin{equation}
 \sigma _{tot}  = 2\int {d^2 b} \left[ {1 - e^{ - n(b,s)/2} }
 \right] .
\end{equation}
 The average number of partonic collisions receives contributions both from hard and
soft physics processes and we write it in the form
\begin{equation}
 n(b,s) = n_{soft} (b,s) + n_{hard} (b,s) ,
\end{equation}
where the soft term parameterizes the contribution of
all the processes for which the partons scatter with
$p_t<p_{tmin}$. It is the only relevant term at  low-energy
and it establishes the overall normalization, while the hard term
is responsible for the high-energy rise. From (\ref{mult}), we
approximate this term with
\begin{equation}
 n_{hard} (b,s) = A(b,s)\sigma _{jet} (s) , \\
 \label{nhard}
 \end{equation}
where the minijet cross section drives the rise due to the
increase of the number of partonic collisions  with the energy and
$A(b,s)$ is the overlap function  which  depends on   the (energy
dependent) spatial distribution of partons inside the colliding
hadrons. In some older models \cite{durand} a simpler factorized
expression for $n(b,s)$ was used,  with  the overlap function
depending  only on $b$. However, when up-to-date realistic parton
densities are used,   such impact parameter distributions,
inspired by  constant hadronic form factors, led to an excessive
rise of $\sigma_{tot}$ with the energy. In our BN model we include an
$s$-dependence in the overlap function that has to tame the strong
growth due to the fast asymptotic rise of $\sigma_{jet}$
\cite{ourmodel}.

We identify soft gluon emissions from the colliding partons as the
physical effect responsible for the attenuation of the rise of the
total cross section. These emissions influence matter distribution
inside of the hadrons, hence changing the overlap function. They
break collinearity between the colliding partons, diminishing the
efficiency of the scattering process. The number of soft emissions
increases with the energy and this makes their contribution
important, also at very high energy. The calculation of
this effect
uses a semiclassical approach based on a Block-Nordsieck inspired
formalism \cite{ddt,*pp,*oldkt}, the basic assumption of this
technique is that all emissions are independent from each other,
so the number of gluons emitted follows a Poisson distribution.
Thereof one obtains a distribution of the colliding partons as
function of the transverse momentum of the soft gluons emitted in
the collision, i.e.
\begin{equation}
d^2P({\bf K_\perp})=d^2{\bf  K_\perp} {{1} \over{(2\pi)^2}}\int
d^2 {\bf b}\ e^{i{\bf K_\perp\cdot b} -h( b )}\ ,
\end{equation}
the factor \ $h(b)$ is given by

\[
h(b) = \int {d^3 } n_g (k)[1 - e^{ - ik_ \bot   \cdot b} ] = \int
{\frac{{d^3 k}}{{2k_0 }}} \sum\limits_{m,n = colors} {|j^{\mu ,m}
(k)j_{\mu ,n} (k)|} [1 - e^{ - ik_ \bot   \cdot b} ],
\]
where $d^3 n_g(k)$ is the distribution for  single gluon emission
in a scattering process and it is linked to the QCD current
$j^{\mu}$ responsible for emission.

\begin{figure}[htb!]
  \begin{center}
  \includegraphics[height=3.9in]{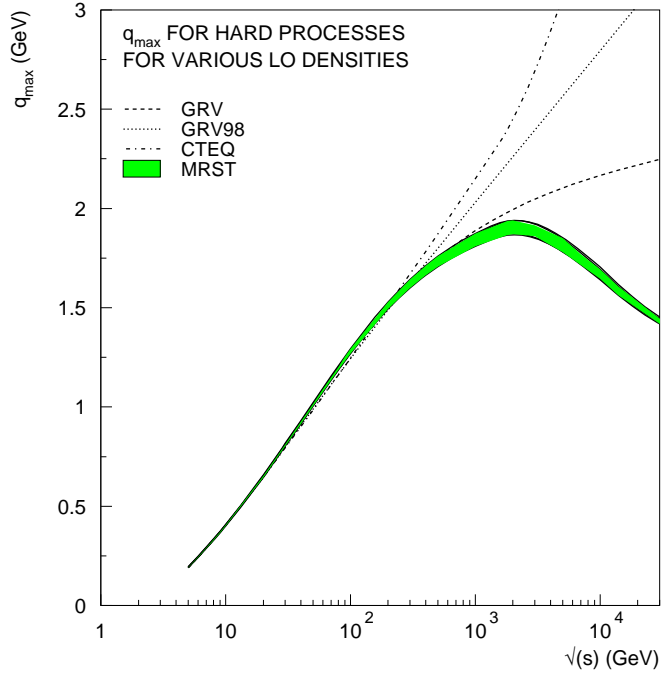}
  \caption{Results for $q_{max}$ using different input parton densities with $p_{tmin}=1.15$ GeV.}
  \label{fig:2}
  \end{center}
\end{figure}

We have proposed to obtain the overlap function as the Fourier
transform of the previous expression of the soft gluon transverse
momentum resummed distribution, namely to put
\begin{equation}
A_{BN}(b,s)=N \int d^2{\bf  K_{\perp}}\  e^{-i{\bf K_\perp\cdot b}}
 {{d^2P({\bf K_\perp})}\over{d^2 {\bf K_\perp}}}={{e^{-h( b,q_{max})}}\over
 {\int d^2{\bf b} \ e^{-h( b,q_{max})}
 }} ,
\end{equation}
with
\begin{equation}
h( b,q_{max}) =  \frac{16}{3}\int_0^{q_{max} }
 {{ \alpha_s(k_t^2) }\over{\pi}}{{d
 k_t}\over{k_t}}\log{{2q_{\max}}\over{k_t}}[1-J_0(k_tb)]
 \label{hbq} ,
\end{equation}
this integral is performed up to a maximum value which is linked
to the maximum transverse momentum allowed by the kinematics for a
single gluon emitted, $q_{max}$ \cite{greco}. In principle, this
parameter and the overlap function should be calculated for each
partonic sub-process, but in the partial factorization of
Eq.(\ref{nhard}) we  use an average value of $q_{max}$ obtained
considering all the sub-processes that can happen for a given
energy of the main hadronic process\cite{ourmodel}:
\begin{equation}
q_{\max } (s)   = \sqrt {\frac{s} {2}}\, \frac{{\sum\limits_{i,j}
{\int {\frac{{dx_1 }} {{x_1 }}\int {\frac{{dx_2 }} {{x_2 }}
\int_{z_{min}}^1 {dz f_i (x_1) f_j (x_2) \sqrt {x_1 x_2 } (1 - z)}
} } } }} {{\sum\limits_{i,j} {\int {\frac{{dx_1 }} {{x_1 }}\int
{\frac{{dx_2}} {{x_2}} \int_{z_{min}}^1 {dz} f_i (x_1)f_j (x_2) }
} } }} ,
\label{qmax}
\end{equation}
with $z_{min}=4p_{tmin}^2/(s x_1 x_2)$. Notice that consistency of
the calculation requires that the  PDF's used in Eq.~ (\ref{qmax})
be the same as those used in $\sigma_{jet}$. In
\figurename{\ref{fig:2}} are presented our results for $q_{max}$
as function of $\sqrt{s}$ using $p_{tmin}=1.15$ GeV.

\begin{figure}[htb!]
  \begin{center}
  \includegraphics[height=3.9in]{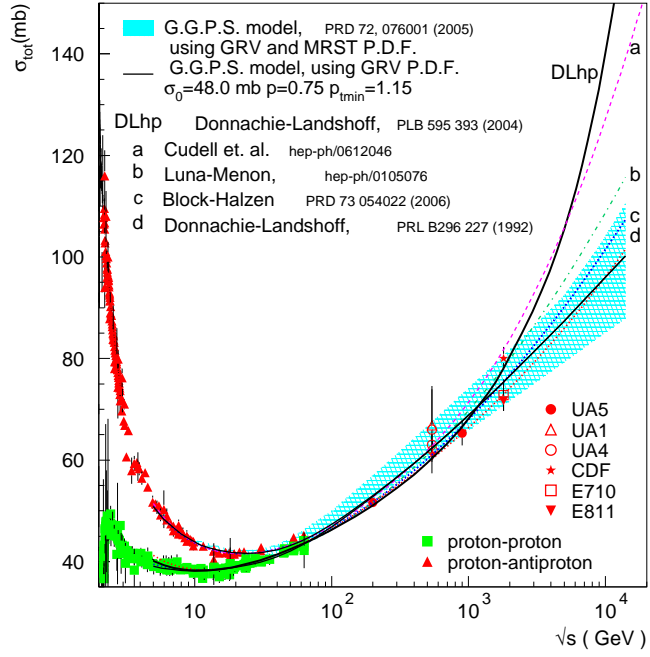}
  \caption{Results from our total cross-section model (for different parton densities) compared with data \protect\cite{data} and with other
  models \protect\cite{bibmodels}.}
  \label{fig:3}
  \end{center}
\end{figure}

The integral in (\ref{hbq}) has another relevant feature, it
extends down to zero momentum values, and to calculate it we have
to take an expression of $\alpha_s$ different from the
perturbative QCD expression which is singular and not integrable
in (\ref{hbq}). We use a phenomenological expression\cite{npb382},
which coincides with the usual QCD limit for large $k_t$, and is
singular but integrable for $k_t \to 0$:
\begin{equation}
\alpha_s(k_t^2)={{12 \pi}\over{33-2N_f}} {{p}
\over{\ln[1+p({{k_t}\over{\Lambda}})^{2p}]}} .
\end{equation}
This expression   for $\alpha_s$ is inspired by the Richardson
expression for a linear confining potential \cite{Richardson}, and
we find  for the parameter $p$ that
\begin{itemize}
\item  $p<1$ to have a convergent integral (unlike
the case of the Richardson potential where $p=1$)
\item $p>1/2$ for
the correct analyticity in the momentum transfer variable.
\end{itemize}
\figurename{\ref{fig:3}} \cite{lastPLB} shows our predictions, obtained 
for the total cross-section  using a set of
phenomenological values for $p_{tmin}$ and $p$, and varying the
parton densities. We also make a comparison with data and other current models.

\section{Restoration of Froissart Bound}
The Froissart Martin Bound \cite{froissmar} states that
$\sigma_{tot}$ cannot rise faster than a function which is
proportional to  $log^2(s)$. In order to see that in our model
this bound is respected, we approximate our total cross section at
very large energies as
\begin{equation}
 \sigma _{tot}  \approx 2\pi \int {db^2 \left[ {1 - e^{ - n_{hard} (b,s)/2} } \right]}
\label{sigtothard} ,
\end{equation}
with $ n_{hard} (b,s) \approx \sigma _{jet} (s)A_{hard} (b,s)$. We then
 take for $\sigma_{jet}$ the asymptotic high energy expression:
\[
\sigma_{jet}=\sigma _1 \left( {\frac{s}{{GeV^2 }}}
\right)^\varepsilon ,
\]
with $\sigma_1=$constant and $\epsilon \sim 0.3-0.4$. Being $
 A_{hard} (b,s) \propto e^{ - h(b,s)}$, we can consider in (\ref{hbq}) the infrared limit $k_t  \to 0$
where the integral receives the dominant contribution. In this
limit we have
\[
\alpha _s (k_t^2 ) \approx \left( {\frac{\Lambda }{{k_t }}}
\right)^{2p} ,
\]
apart from logarithmic terms. Then, with
 $ h(b,s) \propto (b\bar \Lambda )^{2p}$ \cite{ourmodel} (again apart from logarithmic terms), we have
\[
 A_{hard} (b) \propto e^{ - (b\bar \Lambda )^{2p} },
 \]
and from this expression
\[
 n_{hard}  = 2C(s)e^{ - (b\bar \Lambda )^{2p} } ,
 \]
with $C(s) = \frac{A_0 \sigma _1}{2}  \left( {\frac{s}{{GeV^2 }}}
\right)^\varepsilon$. The very high energy limit of Eq.~(\ref{sigtothard}) then gives
\begin{equation}
 \sigma _{tot}  \approx 2\pi \int_0^\infty  {db^2 [1 - e^{ - C(s)e^{ - (b\bar \Lambda )^{2p} } } ] \to \left[ {\varepsilon \ln \left( {\frac{s}{{GeV^2 }}} \right)} \right]^{1/p}
 } .
\end{equation}
The asymptotic growth of $\sigma_{tot}$ in our model depends on
the parameter $\epsilon$ which fixes the asymptotic rise of the
minijet cross section, and on $p$ which modulates the infrared
behavior of $\alpha_s$. Notice that $1/2<p<1$ and thus this approximated result links the
restoration of the Froissart bound in our model  with
the infrared behavior of $\alpha_s$. We can now understand why a
knowledge of the confining phase of the strong interaction is
necessary if we want to restore the finite size of the hadronic
interaction.

 \begin{footnotesize}
 
\end{footnotesize}
\end{document}